\documentclass[twocolumn,aps,pra,superscriptaddress,longbibliography]{revtex4-2}
\usepackage{natbib}
\usepackage[T1]{fontenc}
\usepackage{babel}
\usepackage{float}
\usepackage{physics}
\usepackage{amsmath}
\usepackage{amssymb}
\usepackage{amsfonts}
\usepackage{soul}
\usepackage{xcolor}
\usepackage{bbm}
\usepackage[caption=false,font=normalsize,labelfont=sf,textfont=sf]{subfig} 
\usepackage{graphicx,epstopdf}
\usepackage{url}
\usepackage{textcase}
\usepackage{bm}
\usepackage[unicode=true]{hyperref}
\usepackage{listings}
\makeatletter

\hypersetup{colorlinks=true, linkcolor=blue, filecolor=blue, urlcolor=cyan,}
\makeatletter
\newcommand*\bigcdot{\mathpalette\bigcdot@{.5}}
\newcommand*\bigcdot@[2]{\mathbin{\vcenter{\hbox{\scalebox{#2}{$\m@th#1\bullet$}}}}}
\makeatother

\begin{document}

\title{Maximal quantum entanglement at exceptional points via unitary and thermal dynamics.}
\author{Akhil Kumar}
\affiliation{Department of Physical Sciences, Indian Institute of Science Education and Research Kolkata, West Bengal 741246, India}
\author{Kater W. Murch}
\affiliation{Department of Physics, Washington University, St. Louis, Missouri 63130}
\affiliation{Center for Quantum Sensors, Washington University, St. Louis, Missouri 63130}
\author{Yogesh N Joglekar}
\affiliation{Department of Physics, Indiana University-Purdue University Indianapolis, Indianapolis, Indiana 46202}

\begin{abstract}
Minimal, open quantum systems that are governed by non-Hermitian Hamiltonians have been realized across multiple platforms in the past two years. Here we investigate the dynamics of open systems with Hermitian or anti-Hermitian Hamiltonians, both of which can be implemented in such platforms. For a single system subject to unitary and thermal dynamics in a periodic manner, we show that the corresponding Floquet Hamiltonian has a rich phase diagram with numerous exceptional-point (EP) degeneracy contours. This protocol can be used to realize a quantum Hatano-Nelson model that is characterized by asymmetric tunneling. For one unitary and one thermal qubit, we show that the concurrence is maximized at the EP that is controlled by the strength of Hermitian coupling between them. Surprisingly, the entropy of each qubit is also maximized at the EP. Our results point to the multifarious phenomenology of systems undergoing unitary and thermal dynamics. 
\end{abstract}
\maketitle

\section{Introduction}
\label{sec:intro}

In quantum theory, the dynamics of an isolated system are governed by a Hermitian Hamiltonian that gives rise to a unitary time evolution due to the real eigenvalues and orthogonal eigenvectors. However, no quantum system is truly isolated and open quantum systems are ubiquitous in nature. Traditionally such open systems have been described by a trace-preserving, decoherence-inducing Lindblad equation for the density matrix of the system. In 1998, Bender and co-workers showcased a broad class of non-Hermitian Hamiltonians with purely real spectra for a non-relativistic particle on a line~\cite{Bender1998}. These Hamiltonians are characterized by invariance under combined parity ($\mathcal{P}$) and time-reversal ($\mathcal{T}$) transformations. The ensuing, deep mathematical work~\cite{Mostafazadeh2002,Mostafazadeh2002b,Mostafazadeh2010} on a complex extension of quantum theory~\cite{Bender2002,Bender2007} soon gave way to experiments on classical, optical systems~\cite{ElGanainy2018}. The latter was engendered by the observation that imaginary potentials represent gain or loss~\cite{Ruschhaupt2005,Makris2008,Klaiman2008}, and therefore, $\mathcal{PT}$-symmetric, non-Hermitian Hamiltonians faithfully describe open, classical systems with balanced gain and loss. Over the past decade, classical $\mathcal{PT}$-symmetric systems have been investigated in coupled waveguides~\cite{Rter2010}, fiber loops~\cite{Regensburger2012}, optical resonators~\cite{Chang2014,Hodaei2014}, acoustics~\cite{Zhu2014}, mechanical oscillators~\cite{Bender2013}, and electrical circuits~\cite{Schindler2011,Wang2020}. In the past two years, these studies have been extended into the quantum domain with ultracold atoms~\cite{Li2019}, entangled photons~\cite{Klauck2019}, a single NV center~\cite{Wu2019} and a superconducting qubit~\cite{Naghiloo2019}. 

This intense research, particularly on non-Hermitian systems in the quantum domain, is driven by the unusual nature of their degeneracies~\cite{Kato1995}. A prototypical $\mathcal{PT}$-symmetric Hamiltonian has a purely real spectrum (and non-orthogonal eigenvectors) when the non-Hermiticity is small; this is the $\mathcal{PT}$-symmetric region. With increasing non-Hermiticity, the real eigenvalues develop level attraction~\cite{Klaiman2008}, become degenerate, and then turn into complex-conjugate pairs. The region with complex conjugate eigenvalues is called the $\mathcal{PT}$-broken region~\cite{Joglekar2013}. The transition from $\mathcal{PT}$-symmetric to $\mathcal{PT}$-broken region occurs at an exceptional point (EP) where the corresponding eigenvectors also coalesce. This EP degeneracy is distinct from the diabolic-point (DP) degeneracy of Hermitian Hamiltonians where eigenvalues become degenerate while corresponding eigenvectors continue to remain orthogonal. Since EPs are branch points of Riemann manifolds that represent complex energies, they are responsible for the enhanced sensing and adiabatic mode-switch phenomena~\cite{Miri2019,Wiersig2020}. Their novel properties have intensified the efforts to engineer EP landscapes~\cite{Joglekar2014,Zhong2019,Zhang2019} and higher-order EPs in the classical and quantum domains~\cite{Chen2017,Hodaei2017,Xiao2019,QuirozJurez2019,Bian2020} .

Here, we investigate the dynamics of bipartite, few-level systems that undergo coherent, unitary evolution generated by a Hermitian Hamiltonian $H_1(t)$, or a coherent, non-unitary evolution generated by a purely anti-Hermitian Hamiltonian $H_2(t)$. Note that the dynamics with anti-Hermitian Hamiltonian $H_2$ is equivalent to an imaginary-time or thermal evolution with a Hermitian Hamiltonian $iH_2$. Neither of the two Hamiltonians alone supports EP degeneracies. Surprising, we show that by coupling the two, through temporal modulation (in a single qubit) or spatial interaction (in two qubits), a rich landscape of EPs can be engineered. By characterizing the resultant Floquet Hamiltonian, we show that the system parameters can be tuned to create the classic Hatano-Nelson model with asymmetric hopping~\cite{Hatano1996}. Although it is easy to implement in a classical setting, its quantum realization is challenging, because it requires dissipators that are quite different from the spontaneous emission dissipator. Our results show that coupling unitary and thermal dynamics provides a new avenue to generate $\mathcal{PT}$-symmetric effective Hamiltonians. These Hamiltonians span the entire range from the standard $\mathcal{PT}$ dimer with on-site gain and loss to the Hatano Nelson model with asymmetric hopping in truly quantum platforms. In this study, we use a superconducting transmon circuit as a base model where $H_1(t)$ is implemented by a (detuned) Rabi drive, while $H_2(t)$---the ``pure gain-loss term'' ---corresponds to the post-selected dynamics where quantum jumps are ignored~\cite{Klauck2019,Naghiloo2019}. However, our analysis is applicable to broad range of quantum and semiclassical or purely classical models where the ``levels'' represent bosonic modes, and therefore amplification and depletion is possible. 

It is important to keep in mind two points at the outset. First, we use the standard Dirac inner-product to obtain expectations values and make experimentally observable predictions. Under this convention, a system undergoes non-unitary dynamics in both $\mathcal{PT}$-symmetric and $\mathcal{PT}$-broken phases. We do not use the ``$\mathcal{CPT}$ inner-product''  (positive-definite intertwining operator that is valid only in the $\mathcal{PT}$-symmetric phase), where it generates ``unitary'' evolution under a new definition of the adjoint. We also do not use the ``biorthogonal inner-product'' that gives negative or zero norm states. All experimental evidence to date shows that the nature follows the Dirac inner-product. We also note that the $\mathcal{CPT}$ inner-product cannot treat the system at the EP or the $\mathcal{PT}$-broken phase -- parameter regimes that are routinely accessed in experiments.

Second is the subtle effect of post-selection in open quantum systems~\cite{Naghiloo2019,Klauck2019,QuirozJurez2019}. Post-selection not only ignores quantum trajectories that undergo quantum jumps thereby generating a non-Hermitian dynamics in the Monte Carlo wave-function approach~\cite{Molmer1993}. Additionally, because the trajectories that are kept are precisely ones where the excitations do not leave the system, it renormalizes the instantaneous density matrix for the system of interest. Thus, relative weights of the decaying and non-decaying level in the post-selected manifold of a thermal qubit change with time while the fact that the qubit is always found in one of these two levels -- the density matrix has unit trace -- is guaranteed by post-selection. We encode this experimental reality by evolving the system with a non-Hermitian Hamiltonian and then normalizing the resulting density matrix at every instance of time.

The plan of the paper is as follows. In Sec.~\ref{sec:onequbit}, we consider a single qubit that is subject to either Hermitian $H_1(t)$ or anti-Hermitian $H_2(t)$ Hamiltonians in a periodic manner with period $T$. The long-term dynamics of such a system is governed by the Floquet formalism~\cite{Floquet1883,Oka2019,Joglekar2014,Lee2015}. We present the results for the Floquet $\mathcal{PT}$ phase diagram as a function of the average anti-Hermitian strength and the Floquet modulation frequency $\Omega\equiv 2\pi/T$ and discuss their consequences. In Sec.~\ref{sec:Hfloquet} we show that the corresponding Floquet Hamiltonian interpolates between a traditional $\mathcal{PT}$-symmetric dimer model and the classic Hatano-Nelson model. In Sec.~\ref{sec:twoqubits}, we investigate coherence and entanglement between two qubits, one thermal and one unitary, as a result of the Hermitian interaction between them. We show that with judicious choice of parameters, the system can be driven from a product state to a maximally entangled state, and that the concurrence of the two-qubit system is maximized at the EP. We conclude the paper with a brief discussion in Sec.~\ref{sec:disc}. 

\section{Single qubit with unitary or thermal dynamics}
\label{sec:onequbit}

\begin{figure*}[]
	\centering
	\includegraphics[width=\textwidth]{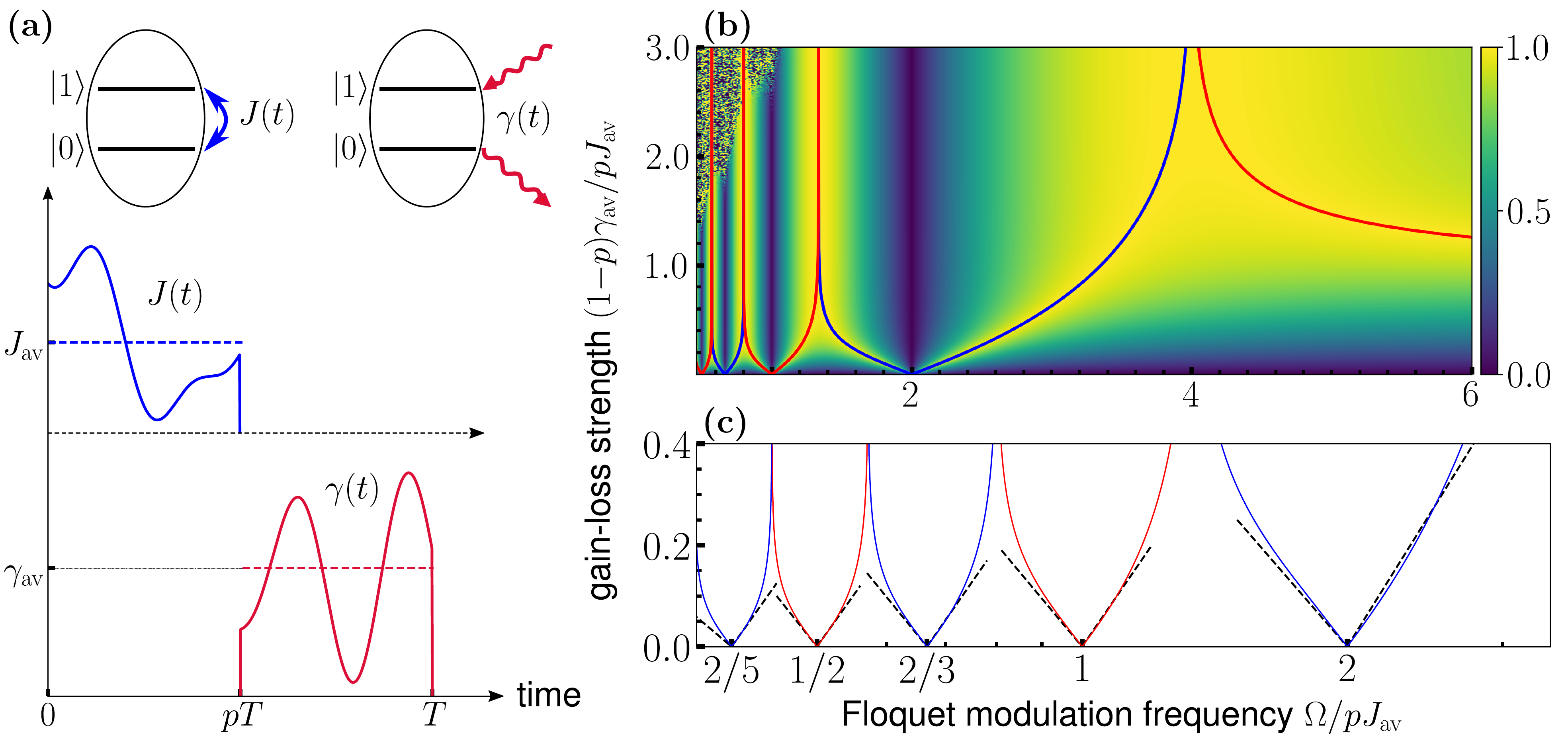} 
	\caption{Single qubit with periodic, Hermitian and anti-Hermitian drives: (a) A two-level system evolves under Rabi drive $J(t)$ for fraction $p$ of the period $T$ and undergoes amplification/depletion with rate $\gamma(t)$ for the remaining time $(1-p)T$. The Floquet dynamics results are solely governed by their temporal averages $J_\mathrm{av}$ and $\gamma_\mathrm{av}$. (b) The heat-map of $I_P(\gamma_\mathrm{av},\Omega)$ shows that EP contours, determined by Eq.(\ref{eq:EPone}) and shown by solid lines, correspond to its maxima. (c) Magnified view of the EP contours in the vicinity of five resonances $\Omega_k=2pJ_\mathrm{av}/k$ for $1\leq k\leq 5$ shows that the linear approximation in Eq.(\ref{eq:slopes})---black, dashed lines---works well at small gain-loss strengths. The EP contours emerging from $\Omega_n/pJ_\mathrm{av}=2/(2n-1)$ are shown in blue and those emerging from $\Omega'_n/pJ_\mathrm{av}=2/(2n)$ are shown in red. Note the logarithmic scale on the horizontal axis in (c).}
	\label{fig:onequbit}
\end{figure*}

Consider a system with a time-periodic Hamiltonian $H(t)=H(t+T)$ that is defined as follows: 
\begin{equation}
	\label{eq:h0}
	H(t)=\left\{\begin{array}{ccc} H_1(t)=H_1^\dagger(t) & & 0\leq t\leq\tau\\
		H_2(t)=-H_2^\dagger(t) & & \tau\leq t\leq T\end{array}\right.
\end{equation}
where $\tau\equiv pT$ and $0\leq p\leq 1$. Thus, within a single period, the system undergoes unitary evolution for time $\tau=pT$ and then an imaginary-time (thermal) evolution for the rest of the time $\beta\equiv(1-p)T$. The corresponding time evolution operators are defined by ($\hbar=1$) 
\begin{align}
	\label{eq:gtau}
	G(\tau)& =\mathbbm{T} e^{-i\int_0^\tau H_1(t')dt'},\\
	\label{eq:gbeta}
	G(\beta)&=\mathbbm{T} e^{-i\int_{\tau}^T H_2(t')dt'},
\end{align}
where $\mathbbm{T}$ stands for the time-ordered product necessary when Hamiltonians at different instances of time do not commute with each other. Since $G(\tau)$ is unitary, its eigenvalues lie on a unit circle and its orthonormal eigenvectors span the space. Although $H_2$ is anti-Hermitian, $G(\beta)=G^\dagger(\beta)$ is Hermitian, with real eigenvalues and complete, orthonormal eigenvectors. Thus, neither $G(\tau)$ nor $G(\beta)$ support EP degeneracies. Since the product of two unitaries is also a unitary, a Hermitian, time-dependent protocol with $G_\mathrm{H}=G(\tau_1)G(\tau_2)$ cannot lead to EP degeneracies. In contrast, a time-evolution operator $G_\mathrm{aH}=G(\beta_1)G(\beta_2)$ with non-Hermitian Hamiltonians may lead to EP degeneracies since the product of two Hermitian matrices $G(\beta_1)$ and $G(\beta_2)$ is not necessarily Hermitian. 

When the system evolves under the Hamiltonian $H(t)$, its long-time dynamics are governed by the time evolution operator for one period, 
\begin{equation}
	\label{eq:gF}
	G_F(T)= G(\beta) G(\tau)\equiv e^{-i TH_F},
\end{equation}
that, in turn also defines the non-Hermitian Floquet Hamiltonian $H_F$. In general, $G_F(T)$ is neither unitary nor Hermitian. Therefore its complex eigenvalues $\lambda_\alpha$ and non-orthogonal (right) eigenvectors $|v_\alpha\rangle$ can be tuned to exhibit EP degeneracies. Equivalently, the Floquet Hamiltonian $H_F$ is neither Hermitian nor anti-Hermitian, and so the fundamental Floquet quasienergies $\epsilon_{\alpha}=+i\ln\lambda_\alpha/T$ are neither real nor purely imaginary. In the unfolded-zone scheme, the Floquet quasienergies are given by $\epsilon_{\alpha n}=\epsilon_\alpha+ n\Omega$~\cite{Floquet1883,Oka2019}. These considerations are valid for general Hamiltonians $H_k(t)$ and thus apply to classical or quantum systems with arbitrary dimensions. 

For a single-qubit case, we start with
\begin{align}
	\label{eq:h1}
	H_{1}(t)&= J(t)\sigma_{x}=\mathcal{PT}H_1(t)\mathcal{PT},\\ 
	\label{eq:h2}
	H_{2}(t)& =i\gamma(t) \sigma_{z}=\mathcal{PT}H_2(t)\mathcal{PT},
\end{align}
where $J(t)$ and $\gamma(t)$ are real, arbitrary functions of time. The schematic for this system is shown in Fig.~\ref{fig:onequbit}(a). The antilinear $\mathcal{PT}$ operator is given by $\mathcal{P}=\sigma_x$ and $\mathcal{T}=*$ (complex conjugation). The time evolution operators are then given by $G(\tau)=\exp(-iJ_\mathrm{av}\tau\sigma_x)$ and $G(\beta)=\exp(+\gamma_\mathrm{av}\beta\sigma_z)$ where 
\begin{align}
	\label{eq:jbar}
	J_\mathrm{av}&=\frac{1}{\tau}\int_0^{\tau} J(t')dt',\\
	\label{eq:gammabar}
	\gamma_\mathrm{av}&=\frac{1}{(T-\tau)}\int_{\tau}^{T}\gamma(t')dt',
\end{align}
denote the temporal averages for the Rabi drive $J(t)$ and the gain-loss strength $\gamma(t)$ respectively. The resulting time evolution operator $G_F(T)=G_0\mathbbm{1}_2+{\bf G}\cdot{\bf\sigma}=G_\mu\sigma_\mu$ can be evaluated explicitly, where ${\bf \sigma}=(\sigma_x,\sigma_y,\sigma_z)$ is a vector with standard Pauli matrices, $\sigma_0=\mathbbm{1}_2$ is the 2-dimensional identity matrix, and sum over the repeated index $\mu\in\{0,x,y,z\}$ is understood. The Floquet eigenvalues are thus given by $\lambda_\pm=G_0\pm |{\bf G}|$ with 
\begin{equation}
	\label{eq:gvec}
	|{\bf G}|=i\sqrt{1-\cosh^2\left[(1-p)T\gamma_\mathrm{av}\right]\cos^2\left(pTJ_\mathrm{av}\right)}.
\end{equation}
It follows that $\lambda_\pm(p,T,\gamma_\mathrm{av},J_\mathrm{av})$ have equal magnitudes when $|{\bf G}|$ is purely imaginary. This defines the $\mathcal{PT}$-symmetric phase of the Floquet Hamiltonian $H_F$. On the other hand, when $|{\bf G}|$ is purely real, the eigenvalues $\lambda_\pm$ have different magnitudes, indicating a $\mathcal{PT}$-broken phase for the system. The EP contours that separate the $\mathcal{PT}$-symmetric region from the $\mathcal{PT}$-broken region are given by $|{\bf G}|=0$, i.e. 
\begin{equation}
	\label{eq:EPone}
	\cos\left(pTJ_\mathrm{av}\right)\cosh\left[(1-p)T\gamma^\mathrm{EP}_\mathrm{av}\right]=\pm 1.
\end{equation}
It is also straightforward to show that the Dirac inner product of Floquet eigenvectors $|v_\pm\rangle$ of $G_F(T)$ is given by $I_P(\gamma_\mathrm{av},\Omega)\equiv|\langle v_+|v_-\rangle|=\min(r,1/r)$ where 
\begin{equation}
	\label{eq:ipone}
	r=\frac{|G_x|}{\left[G_y^2+G_z^2\right]^{1/2}}=\left|\frac{\sin(pTJ_\mathrm{av})}{\tanh\left[(1-p)T\gamma_\mathrm{av}\right]}\right|.
\end{equation}
The Floquet $\mathcal{PT}$-phase diagram and resultant EP contours are characterized solely in terms of the average Rabi drive, Eq.(\ref{eq:jbar}), and average gain-loss strength, Eq.(\ref{eq:gammabar}). They are, therefore, independent of the exact functional forms and make our results widely applicable. Without loss of generality, we take $J_\mathrm{av}>0$ and confine our attention to $\gamma_\mathrm{av}>0$ since the results for negative $\gamma_\mathrm{av}$ are obtained from it by symmetry transformations. 

Figure~\ref{fig:onequbit}(b) shows numerically obtained heat-map of $I_P$ as a function of dimensionless loss strength $(1-p)\gamma_\mathrm{av}/pJ_\mathrm{av}$ and the Floquet modulation frequency $\Omega/pJ_\mathrm{av}$. (We use the frequency $\Omega=2\pi/T$ instead of the modulation period $T$ for ease of comparison with the literature.) Superimposed on the heat map are EP contours obtained via Eq.(\ref{eq:EPone}) with red corresponding to value +1 and blue corresponding to value $-1$. The blue contours emerge from modulation frequencies $\Omega/pJ_\mathrm{av}=\{2,2/3,2/5,\ldots\}$ and the red contours emerge from $\Omega/pJ_\mathrm{av}=\{2/2,2/4,2/6,\ldots\}$. It follows from Eq.(\ref{eq:EPone}) that the $\mathcal{PT}$-broken phase occurs at vanishingly small values of $\gamma_\mathrm{av}$ in the neighborhood of resonances given by $\Omega_k=2pJ_\mathrm{av}/k$ where $k\geq 1$ is an integer~\cite{Joglekar2014,Lee2015,Li2019,LenMontiel2018}. It is worth pointing out that the resonances $\Omega_k$ along the $\gamma_\mathrm{av}=0$ axis are DP degeneracies that terminate the EP contours in the $\gamma_\mathrm{av}>0$ plane. A perturbative expansion in the neighborhood of the $k^\mathrm{th}$ resonance shows that the equation for EP lines that emerge from $\Omega_k$ with equal and opposite slopes is given by 
\begin{equation}
	\label{eq:slopes}
	\gamma^\mathrm{EP}_\mathrm{av}(\Delta\Omega_k)=\pm \frac{k}{2(1-p)}\Delta\Omega_k,
\end{equation}
where $\Delta\Omega_k=\Omega-\Omega_k$. Figure~\ref{fig:onequbit}(c) shows the exact EP contours, Eq.(\ref{eq:EPone}), (red and blue lines) and linear approximation, Eq.(\ref{eq:slopes}), (black dashed lines) in the vicinity of first five resonances. The linear approximation is valid in a region that algebraically shrinks with increasing resonance index $k$, i.e. $|\Delta\Omega_k|\ll (2/k)pJ_\mathrm{av}$. We also note that the inner product vanishes at resonances $\Omega_k=2pJ_\mathrm{av}/k$, indicating Dirac-orthogonal eigenvectors $|v_\pm\rangle$. At the resonance $\Omega=\Omega_k$, the unitary time evolution is trivial, i.e. $G(\tau)=(-1)^k\mathbbm{1}_2$, and therefore the one-period time-evolution operator $G_F(T)$ is purely generated by an anti-Hermitian Hamiltonian. 

\begin{figure*}[]
	\centering
	\includegraphics[width=\textwidth]{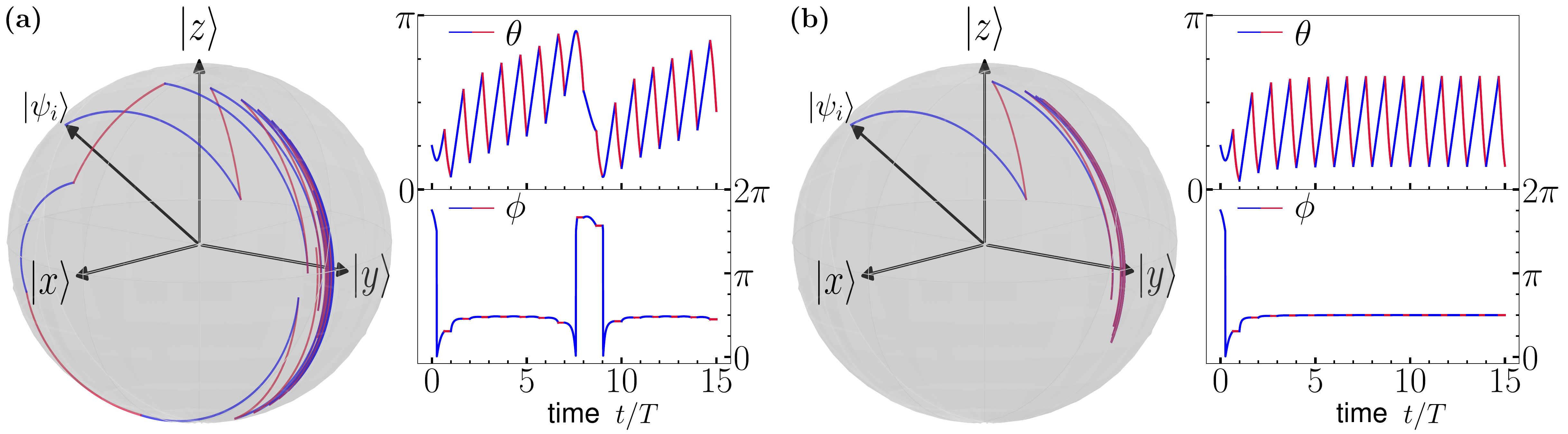} 
	\caption{Qubit trajectory under unitary (blue lines) and thermal (red lines) evolution for a qubit initialized in the state $|\psi(0)\rangle= \left(\ket{+x} + \ket{-y} + \ket{+z} \right)/\sqrt{3}$. Here $|\pm x\rangle$ ($|\pm y\rangle, |\pm z\rangle$) stand for the eigenstates of $\sigma_x$ ($\sigma_y, \sigma_z$) with eigenvalues $\pm 1$ respectively. (a) When $(1-p)\gamma_\mathrm{av}/pJ_\mathrm{av}=1.0$ and $\Omega/pJ_\mathrm{av}=2.5\pi$, the system is in the $\mathcal{PT}$-symmetric state. Starting from $|\psi(0)\rangle$, the system traces a periodic trajectory consisting of precession about the x-axis (unitary dynamics) and motion towards the north pole along meridians (thermal dynamics). Corresponding evolution of the polar angle $\theta(t)$ and azimuthal angle $\phi(t)$ on the Bloch sphere as a function of time measured in the units of $T$ is also shown. (b) When $(1-p)\gamma_\mathrm{av}/pJ_\mathrm{av}=1.25$ and $\Omega/pJ_\mathrm{av}=2.5\pi$, the system is in the $\mathcal{PT}$-broken phase. As a result, the same initial state $|\psi(0)\rangle$ reaches a stroboscopic steady-state in the y-z plane, while the micromotion generates fast oscillations in the polar angle $\theta(t)$.}
	\label{fig:bloch}
\end{figure*}

It also follows from Eq.(\ref{eq:EPone}) that when modulation frequency is close to the node of cosine function, i.e. $\Omega\approx\Omega'_k\equiv 2pJ_\mathrm{av}/(k+1/2)$, the $\mathcal{PT}$-symmetric phase extends to arbitrarily large values of gain-loss strengths. A perturbative expansion in the vicinity of the node, $\Omega\approx\Omega'_k$, shows that for $(1-p)\gamma_\mathrm{av}/pJ_\mathrm{av}\gg 1$, the EP contour is characterized by 
\begin{equation}
	\label{eq:nodes}
	\gamma^\mathrm{EP}_\mathrm{av}(\Delta\Omega_k)=-\frac{pJ_\mathrm{av}}{\pi(k+1/2)(1-p)}\ln\left[\frac{\pi\Delta\Omega'_k}{pJ_\mathrm{av}}\right],
\end{equation}
where $\Delta\Omega'_k=|\Omega-\Omega'_k|$. Lastly, the high-frequency limit of the EP contours in Fig.~\ref{fig:onequbit}b is obtained most easily from Eq.(\ref{eq:ipone}) where the constraint $r=1$ in the limit $T\rightarrow 0$ gives
\begin{equation}
	\label{eq:highfreq}
	\gamma^\mathrm{EP}_\mathrm{av}(\Omega\rightarrow\infty)=\frac{pJ_\mathrm{av}}{(1-p)}.
\end{equation}
This result is expected because in the high-frequency limit, the Floquet problem is equivalent to a static problem with time-averaged Rabi drive $pJ_\mathrm{av}$ and gain-loss strength $(1-p)\gamma_\mathrm{av}$. 

To understand the emergence of $\mathcal{PT}$-symmetric and $\mathcal{PT}$-broken phases in the Floquet dynamics, we numerically investigate the evolution of a state $|\psi(0)\rangle$ on the Bloch sphere under the influence of Hamiltonian $H(t)$. Under the unitary dynamics introduced by $J(t)\sigma_x$, the state traces (part of a) circle in the y-z plane, whereas under the thermal dynamics introduced by $i\gamma(t)\sigma_z$, it travels along a meridian toward the north pole. We empahsize that the post-selection process ensures that the qubit stays on the Bloch sphere surface. Figure~\ref{fig:bloch}(a) shows the state's trajectory during unitary (blue) and thermal (red) evolutions when the system is in the $\mathcal{PT}$-symmetric phase. We see that the Bloch angles $\theta(t)$ and $\phi(t)$ undergo fast oscillatory behavior (micromotion), whereas the $\mathcal{PT}$-symmetric phase is signified by a periodic behavior over timescale of several periods. Corresponding results for a system in the $\mathcal{PT}$-broken phase are shown in panel (b). Starting with the same initial state, the qubit trajectory stabilizes in the y-z plane, with rapid (micromotion) oscillations in the polar angle $\theta(t)$ and stroboscopic steady-state behavior. The results presented in Figs.~\ref{fig:bloch}(a)-(b) are representative, and show that the $\mathcal{PT}$-symmetric or $\mathcal{PT}$-broken regions emerges due to competition between precession around the x-axis in the unitary part of the dynamics, and travel along meridians in the thermal part of the dynamics. 

In contrast to the traditionally studied models of periodic Hamiltonian or Lindblad dynamics, our model leverages alternating unitary and thermal evolutions to create EP contours that, in simple cases, can be analytically determined for arbitrary $J(t)$ and $\gamma(t)$. That, in turn, allows us to obtain the exact Floquet Hamiltonian $H_F$. In the following section, we will investigate its the behavior. 

\section{Effective Floquet Hamiltonian}
\label{sec:Hfloquet}

The Floquet Hamiltonian defined by Eq.(\ref{eq:gF}) is the static Hamiltonian that generates the same time evolution as the time-periodic Hamiltonian $H(t)$ only at stroboscopic times, i.e. $t_n=nT$. At intermediate times $t_{n-1}<t< t_n$, the micromotion generated by $H(t)$ does not match the dynamics generated by $H_F$. However, since the micromotion contribution is periodic in time, the long-term dynamics are purely determined by the Floquet Hamiltonian, that, in the frequency basis, is given by $H_F=H-i\partial_t$~\cite{Joglekar2014,Lee2015}. In principle, it is possible to obtain the Floquet Hamiltonian over the entire parameter space as $H_F=+i\ln G_F(T)/T$. By parameterizing it as $H_F=h_0\mathbbm{1}_2+{\bf h}\cdot{\bf\sigma}=h_\nu\sigma_\nu$ gives $\exp(-ih_0T)=\pm1$ and the following implicit equations for the vector ${\bf h}$, 
\begin{align}
	\label{eq:hxx}
	\frac{h_x}{|{\bf h}|}& =\frac{\tan\left(pTJ_\mathrm{av}\right)}{\tan(|{\bf h}|T)},\\
	\label{eq:hzz}
	\frac{h_z}{|{\bf h}|}& =i\frac{\tanh\left[(1-p)T\gamma_\mathrm{av}\right]}{\tan(|{\bf h}|T)},\\
	\label{eq:hyy}
	\frac{h_y}{|{\bf h}|}& =i\frac{\tan\left(pTJ_\mathrm{av}\right)\tanh\left[(1-p)T\gamma_\mathrm{av}\right]}{\tan(|{\bf h}|T)}.
\end{align}
The constraint of antilinear ($\mathcal{PT})$ symmetry on the Floquet Hamiltonian implies that the components of ${\bf h}$ are real or imaginary, but not complex. Thus, ${\bf h}\cdot{\bf h}$ is a real (positive, zero, or negative) quantity, and $|{\bf h}|=0$ characterize the EP degeneracies of the Floquet Hamiltonian. 

One remarkable feature of our model is that $H_F$ represents a ``two-site'' system with {\it asymmetric tunneling as well as on-site gain and loss}. It is a combination of the Hatano-Nelson model ($H_\mathrm{HN}=A\sigma_x+iB\sigma_y$) and the $\mathcal{PT}$-symmetric dimer ($H_\mathrm{dimer}=A\sigma_x+iB\sigma_z$). The ratio of these contributions is given by $h_y/h_z=\tan(pJ_\mathrm{av}T)$ and thus can be arbitrarily varied. Specifically, at $\Omega_n/pJ_\mathrm{av}=2/n$, the $h_z$-term dominates and gives an anti-Hermitian Hamiltonian. On the other hand, when $\Omega'_k/pJ_\mathrm{av}=2/(k+1/2)$, the $h_y$-term dominates, and gives rise to the Hatano-Nelson model with a divergent $\mathcal{PT}$-symmetry breaking threshold; see Fig.~\ref{fig:onequbit}(b). 

Further insight is gained when we move along the EP contours defined by Eq.(\ref{eq:EPone}) or equivalently $|{\bf h}|=0$. Since the Hamiltonian satisfies the equation $(H_F-h_0\mathbbm{1}_2)^2=0$, the power series expansion for the time-evolution operator terminates at first order, i.e. $G_F(T)=\pm\left[\mathbbm{1}_2-i{\bf h}\cdot{\bf\sigma}T\right]$. Therefore, the Floquet Hamiltonian simplifies to
\begin{align}
	\label{eq:hx}
	h_x(\gamma_\mathrm{av})& =\frac{1}{T}\sinh[(1-p)\gamma_\mathrm{av}T]=-h_x(-\gamma_\mathrm{av}), \\
	\label{eq:hy}
	h_y(\gamma_\mathrm{av})&=\frac{i}{T}\tanh[(1-p)\gamma_\mathrm{av}T]=-h_y(-\gamma_\mathrm{av}), \\
	\label{eq:hz}
	h_z(\gamma_\mathrm{av})& =Th_xh_y=+h_z(-\gamma_\mathrm{av}).
\end{align}

Another remarkable feature of the Floquet model is that the DP degeneracies at $\gamma_\mathrm{av}=0, \Omega=\Omega_k$ are continuously connected to the EP degeneracies of the non-Hermitian Floquet problem. At any point along the EP contour, the norm of a generic state $|\psi(t_n)\rangle=G_F(T)^n|\psi(0)\rangle$ grows quadratically with time $t_n$; however, as one approaches the DP degeneracies on the $\gamma_\mathrm{av}=0$ axis, the evolution is unitary and the norm is preserved. This ``proximity to the DP''~\cite{roberto2021ondemand} can be characterized by eigenvalues $\Lambda_\pm$ of the Hermitian, positive-definite operator $G_F^\dagger(T)G_F(T)$. They are given by 
\begin{equation}
	\label{eq:gdaggerg}
	\Lambda_\pm =\exp\left[\pm 2(1-p)T\gamma_\mathrm{av}\right].
\end{equation}
Evidently, as one approaches a DP degeneracy along the $\Omega$ axis, $\Lambda_\pm\rightarrow 1$ as is expected for a unitary time evolution. 

\section{Entanglement between thermal and unitary qubits}
\label{sec:twoqubits}

\begin{figure*}[]
	\centering
	\includegraphics[width=\textwidth]{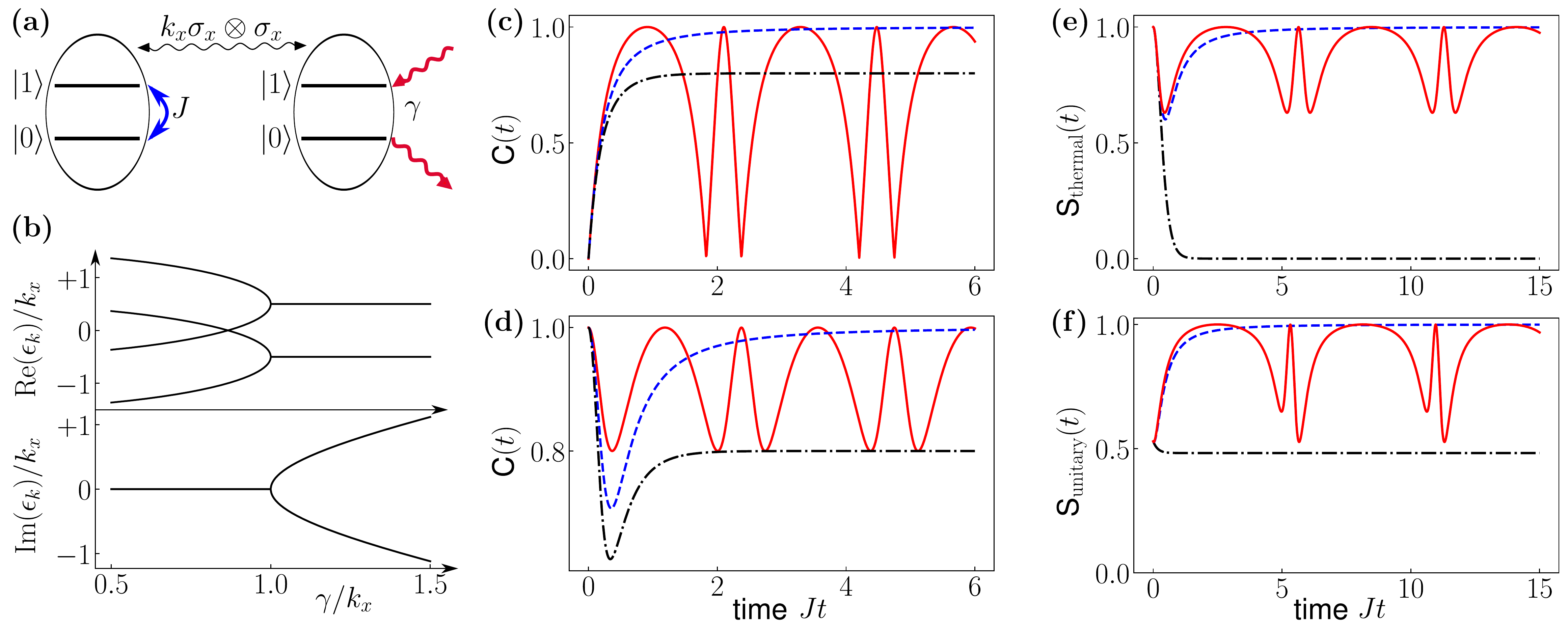} 
	\caption{Two coupled qubits. (a) A unitary qubit with Rabi drive $J\sigma_x$ is coupled to a thermal qubit with Hamiltonian $i\gamma\sigma_z$ by a Hermitian interaction with strength $k_x$. When uncoupled, the system is in the $\mathcal{PT}$-broken phase due to the thermal qubit. (b) Eigenvalues $\epsilon_k$ of $H_2$ as a function of $\gamma/k_x$ show a second-order EP at $\gamma=k_x$. (c) With initial state $|00\rangle$, the concurrence $C(t)$ is periodic when $\gamma/k_x=0.75$ (red solid), saturates to its maximum value of unity at the EP (blue dashed), and is suppressed from its maximum at $\gamma=1.25k_x$ (black dot-dashed). (d) Starting from a Bell state, $|\psi_2(0)\rangle=(|00\rangle+|11\rangle)/\sqrt{2}$, similar behavior is observed. For (c)-(d), $k_x/J=2$. (e) Starting from a maximally-mixed two-qubit state, $S_\mathrm{t}(t)$ reaches zero when $k_x=0$ (black dot-dashed). With increasing $k_x$, $S_\mathrm{t}(t)$ reaches maximum at the EP, $k_x=\gamma=1.5J$ (blue solid), and then becomes oscillatory when $k_x=1.6J$ (red solid). (f) The unitary qubit entropy $S_\mathrm{u}(t)$ also shows clear signature of the EP at $k_x=\gamma=1.5J$. The normalized time is measured in units of $Jt$.}
	\label{fig:twoqubit}
\end{figure*}

In previous sections, we have considered a single qubit that undergoes unitary and thermal evolutions in a periodic manner. In this section, we consider the dynamics of two qubits, one unitary and one thermal, that are coupled by a Hermitian interaction, Fig.~\ref{fig:twoqubit}(a). When the qubits do not interact, the unitary qubit has a constant entropy and its density matrix undergoes precession within the Bloch sphere at a constant radius. In contrast, the thermal qubit's entropy decreases as its density matrix moves along a meridian radially outward and toward the north pole on the Bloch sphere. In the presence of interaction, the two qubit Hamiltonian $H_2$ is given by 
\begin{align}
	\label{eq:h2qubit}
	H_2 &= J \mathbbm{1}_{2} \otimes \sigma_{x} +i\gamma \sigma_z \otimes \mathbbm{1}_{2} + k_{x} \sigma_{x} \otimes \sigma_{x},\\
	\label{eq:eps2}
	\epsilon_k &=\pm J\pm\sqrt{k_x^2-\gamma^2}=\pm J\pm\Delta,
\end{align}
where $\epsilon_k$ denote its four eigenvalues. The Hamiltonian $H_2$ is commutes with the $\mathcal{PT}$-operator where $\mathcal{P}=\sigma_x\otimes\sigma_x$ and $\mathcal{T}$ is given by complex conjugation. It follows that the eigenvalues are complex for $k_x\leq\gamma$ with a second-order EP at $\gamma=k_x$. The $\mathcal{PT}$-symmetric phase with purely real eigenvalues emerges when the Hermitian coupling strength $k_x$ exceeds $\gamma$. Figure~\ref{fig:twoqubit}(b) shows the real and imaginary parts of the eigenvalues $\epsilon_k$ as a function of non-Hermiticity $\gamma$ for a Rabi-drive strength $J=0.5k_x$.  

The dynamics of the two qubit system is given by 
\begin{equation} 
	\label{eq:rho2}
	\rho_2(t)=\frac{G_2(t)\rho_2(0)G^\dagger_2(t)}{\mathrm{Tr}\left[G_2(t)\rho_2(0)G^\dagger_2(t)\right]}
\end{equation}
where $\rho_2(t)$ is the normalized two-qubit density matrix and $G_2(t)=\exp(-iH_2 t)$ is the non-unitary time evolution operator. This normalization reflects the experimental reality that post-selection only keeps quantum trajectories where, by ignoring quantum jumps, the trace of the two-particle density matrix is maintained. We emphasize that this normalization occurs in both $\mathcal{PT}$-symmetric or $\mathcal{PT}$-broken phases, and is not related to the norm-conserving $\mathcal{CPT}$ evolution for two qubits in the $\mathcal{PT}$-symmetric phase~\cite{Bhosale2021}. To investigate the entanglement between the two qubits, we calculate the Wootter's concurrence~\cite{Wootters1998,Wootters2001} $C(t)=\max\{0,c_1-c_2-c_3-c_4\}$ where $c_k(t)$ are the eigenvalues, in decreasing order, of the positive-semidefinite matrix 
\begin{equation}
	\label{eq:conc}
	R(t)=\left[\rho_2(t)(\sigma_y\otimes\sigma_y)\rho_2^*(t)(\sigma_y\otimes\sigma_y)\right]^{1/2}.
\end{equation}
We also obtain the individual qubit entropies $S_\mathrm{u,t}=\mathrm{Tr}\left[\rho_\mathrm{u,t}\log_2\rho_\mathrm{u,t}\right]$ where $\rho_\mathrm{u}(t)=\mathrm{Tr_t}\rho_2(t)$ is the reduced density matrix for the unitary qubit and $\rho_\mathrm{t}(t)=\mathrm{Tr_u}\rho_2(t)$ is the reduced density matrix for the thermal qubit. 

Figure~\ref{fig:twoqubit}(c) shows the development of concurrence between two qubits that are initialized in respective ground states, i.e. $|\psi_2(0)\rangle=|00\rangle$ and are strongly coupled, i.e.  $k_x=2J$. When $\gamma/k_x=0.75$ (red solid), the system is in the $\mathcal{PT}$-symmetric phase and $C(t)$ shows a periodic behavior that reaches unity, thereby indicating maximally entangled qubits. At the EP $\gamma=k_x$ (blue dashed), the $C(t)$ saturates to the maximum value of unity, thereby indicating that the two qubits approach a maximally entangled steady state. In the $\mathcal{PT}$-broken regime, $\gamma=1.25k_x$ (black dot-dashed), the concurrence reaches a lower steady-state value indicating reduced entanglement. Corresponding results for a system starting in an initial state $|\psi_2(0)\rangle=(|00\rangle+|11\rangle)/\sqrt{2}$ are shown in Fig.~\ref{fig:twoqubit}(d). Once again, we find the concurrence reaches steady-state value of unity at the EP (blue dashed). 
	
These results can be analytically obtained in certain parameter ranges. For two-qubit, pure states with time-dependent, normalized coefficients $|\phi_2(t)\rangle=a_{00}(t) |00\rangle+a_{01}(t)|01\rangle+a_{10}(t)|10\rangle+a_{11}(t)|11\rangle$, it is straightforward to show~\cite{Wootters2001} that the concurrence is given by $C(t)=2|a_{00}a_{11}-a_{10}a_{01}|$. We first obtain the non-unitary time-evolution operator as  
\begin{align}
\label{eq:g2}
G_2(t)&=\frac{\sin(\Delta t)}{\Delta}\left[\begin{array}{cccc}
{\mathcal C}A_+ & -i{\mathcal S}A_+ & -k_x\mathcal{S} & -ik_x{\mathcal C}\\
-i{\mathcal S}A_+ & {\mathcal C}A_+ & -i k_x{\mathcal C} & -k_x{\mathcal S}\\
-k_x{\mathcal S} & -ik_x {\mathcal C} & {\mathcal C}A_{-} & -i{\mathcal S}A_{-}\\
-ik_x{\mathcal C}& -k_x{\mathcal S} & -i{\mathcal S}A_{-} & {\mathcal C}A_{-}
\end{array}\right]
\end{align}
where ${\mathcal C}=\cos(Jt)$, $\mathcal{S}=\sin(Jt)$, and $A_\pm=\Delta\cot(\Delta t)\pm\gamma$. Note that the diverging nature of $A_\pm$ in the limit $t\rightarrow 0$ is compensated by the vanishing prefactor $\sin(\Delta t)$ in Eq.(\ref{eq:g2}), and gives $G_2(t\rightarrow 0)=\mathbbm{1}_4$. In the $\mathcal{PT}$-symmetric phase $\Delta=\sqrt{k_x^2-\gamma^2}$ is real and gives rise to the oscillatory behavior.  In the $\mathcal{PT}$-broken phase it is purely imaginary, and gives rise to steady-state values for the normalized density matrix. The time-evolution operator $G_2(t)$ remains continuous across the $\mathcal{PT}$ transition, and thus, the results at the EP ($\Delta=0$) are obtained by taking the limit $\Delta\rightarrow 0$ of the expressions in Eq.(\ref{eq:g2}).

Using the time-evolution operator, it is straightforward to show that the time-dependence concurrence for an initial state $|\psi_2(0)\rangle=|00\rangle$ is given by 
\begin{equation}
\label{eq:c00s}
C(t)=\left|\frac{2k_x \left[\Delta\cot(\Delta t)+\gamma\right]}{k_x^2+\left[\Delta\cot(\Delta t)+\gamma\right]^2}\right|\leq 1.
\end{equation}
It follows that when $\Delta>0$ is real, the initially uncorrelated qubits become periodically maximally entangled, $C=1$, with period given by $T_\Delta=\pi/(2\Delta)$. It also reaches zero when $A_{+}=\Delta\cot(\Delta t)+\gamma$ is either zero or diverges. At the exceptional point $\gamma=k_x$, the  concurrence is given by 
\begin{equation}
\label{eq:c00e}
C(t)=\left|\frac{2\gamma t(1+\gamma t)}{ (\gamma t)^2+(1+\gamma t)^2}\right|
\end{equation}
and approaches unity at times $\gamma t\gg 1$. When the system is in the $\mathcal{PT}$-broken phase, $\Delta$ is purely imaginary and the term $\Delta\cot(\Delta t)\rightarrow |\Delta|$ at long times. Therefore, when $\gamma>k_x$, the long-time, steady-state concurrence is suppressed from its unity value at the EP and is given by $C_\mathrm{ss}=k_x/\gamma<1$. Similar analysis for a symmetric Bell state as initial state gives the results in Fig.~\ref{fig:twoqubit}(d). 

Lastly, we consider the dynamics of the purity of each qubit as a function of the coupling $k_x$ between them. When the two qubits are initially in the maximally mixed state, i.e. $\rho_2(0)=\mathbbm{1}_4/4$,   Fig.~\ref{fig:twoqubit}(e) shows that, at $k_x=0$, (black dot-dashed) the thermal qubit entropy $S_\mathrm{t}(t)$ starts at unity and decreases to zero at long times. This is becuase the thermal qubit approaches the pure state at the north pole, thereby monotonically decreasing its entropy. With increasing $k_x\leq\gamma$, $S_\mathrm{t}(t)$ decreases and then reaches a steady-state value that rises to unity at the exceptional point $k_x=\gamma=1.5J$ (blue dashed). Past the EP, when $k_x=1.6J>\gamma$, the entropy shows oscillations bounded between 0 and 1 that are characteristic of the $\mathcal{PT}$-symmetric phase (red solid). The reduced density matrix of the unitary qubit, on the other hand, remains maximally mixed irrespective of $k_x$ value and therefore $S_\mathrm{u}(t)=1$. Figure~\ref{fig:twoqubit}(f) shows the entropy $S_\mathrm{u}(t)$ for an initial two-qubit state $\rho_2(0)=0.25\mathbbm{1}_4+0.19\mathbbm{1}_2\otimes\sigma_z+0.23\sigma_z\otimes\mathbbm{1}_2+0.19\sigma_z\otimes\sigma_z$. When $k_x=0$ (black dot-dashed), the entropy of the unitary qubit remains constant with time as is expected. As $k_x$ increases, $S_\mathrm{u}(t)$ rises and reaches unity at the EP, $k_x=\gamma=1.5J$ (blue dashed). For $k_x=1.6J>\gamma$, it shows oscillatory behavior (red solid). These trends, too, can be analytically understood by using the exact time-evolution operator, Eq.(\ref{eq:g2}). It is interesting that the entropy of the interacting unitary qubit is always higher than its constant value in the absence of interaction with the thermal qubit. 

\section{Discussion}
\label{sec:disc}
Recent progress on non-Hermitian, minimal quantum systems~\cite{Li2019,Wu2019,Klauck2019,Naghiloo2019} and unitary quantum simulators that can implement non-unitary and Lindblad dynamics~\cite{Sparrow2018,Lin2021} makes the interplay between unitary and thermal dynamics possible. In this article, we investigated this interplay through two models. For a single qubit that periodically undergoes thermal and unitary evolutions, we have shown that the resultant dynamics has a rich Floquet structure with numerous EP contours that separate $\mathcal{PT}$-symmetric and $\mathcal{PT}$-broken regions. By characterizing the corresponding Floquet Hamiltonian $H_F$, we have shown that this protocol can be used to realize the classic Hatano-Nelson model in a minimal, quantum platform. Although the Hamiltonian $H_\mathrm{HN}$ is unitarily equivalent to the $\mathcal{PT}$-dimer Hamiltonian, the latter is easily realized through post-selection in the presence of spontaneous-emission dissipator~\cite{Naghiloo2019}. In contrast, realizing the former has been challenging since it requires bath-engineering new dissipators that, after post-selection, will give rise to a $\sigma_y$-term in the anti-Hermitian part of the effective Hamiltonian. In the second model, we have studied the restoration of $\mathcal{PT}$-symmetry, indicated by a periodic behavior, that emerges when a thermal qubit is moderately or strongly coupled to a unitary qubit. In this case, we have shown that the entanglement between the two qubits is maximized at the EP, just as the each qubit shows maximal entropy as well.  

The analysis presented here applies to higher-dimensional systems as well. When higher-dimensional representations of SU(2)---qutrits, qudits---are considered, results in Sec.~\ref{sec:onequbit} remain valid with higher-order EP contours~\cite{QuirozJurez2019,Bian2020}. They also remain qualitatively same for other Hermitian couplings $K_{\mu\nu}\sigma_\mu\otimes\sigma_\nu$ between the unitary and thermal qubits. It is experimentally more challenging to construct bipartite systems where one undergoes unitary evolution while the second has only thermal dynamics. However, our results show that such systems offer entangled steady-states, something that is not possible in purely Hermitian dynamics.  A complete characterization of $H_F$ generated from the interplay of thermal and unitary dynamics will enable the engineering of arbitrary non-Hermitian Hamiltonians in truly quantum systems.


\section*{acknowledgment}
This work was supported, in part, by ONR Grant No. N00014-21-1-2630. 


\bibliography{Paper.bib}

\end{document}